%% file: a.tex
\documentclass{article}
\usepackage[preprint]{spconf}

\usepackage{cite}
\usepackage{amsmath,amssymb,amsfonts}
\usepackage[pdftex]{graphicx}
\graphicspath{{pdf/}}
\DeclareGraphicsExtensions{.pdf,.jpeg,.png}
\usepackage{amsthm}
\usepackage{booktabs}
\usepackage{siunitx}
\usepackage{algorithm}

\usepackage[caption=false,font=footnotesize]{subfig}
\usepackage{hyperref}
\hypersetup{colorlinks=true,linkcolor=black,citecolor=black,urlcolor=blue}

\input{skroty}

\begin{document}
%
\title{Super-resolution of periodic signals from short sequences of samples}
%
%
%

\name{Marek W. Rupniewski}
\address{Institute of Electronic Systems\\
	Warsaw University of Technology,\\
	Nowowiejska 15/19, 00-665 Warsaw, Poland\\
  Email: Marek.Rupniewski@pw.edu.pl%
}%

\maketitle
\copyrightnotice{\begin{minipage}{\textwidth}\centering\small
        \copyright 2021 IEEE.  Personal use of this material is permitted.  Permission from IEEE must be obtained for all other uses, in any current or future media, including reprinting/republishing this material for advertising or promotional purposes, creating new collective works, for resale or redistribution to servers or lists, or reuse of any copyrighted component of this work in other works.
\end{minipage}}
\toappear{\begin{minipage}{\textwidth}\small M. W. Rupniewski, "Super-Resolution Of Periodic Signals From Short Sequences Of Samples," \emph{ICASSP 2021 - 2021 IEEE International Conference on Acoustics, Speech and Signal Processing (ICASSP)}, 2021, pp. 4995-4999, doi: \href{https://doi.org/10.1109/ICASSP39728.2021.9413593}{10.1109/ICASSP39728.2021.9413593}.
\end{minipage}}

\begin{abstract}
  Reconstruction of undersampled periodic signals of unknown period is an
  important signal processing operation. It is especially difficult operation
  when the sequences of samples are short and no information on the inter-sequence
  time distances is given. For such a case, there exist some algorithms that allow for approximation 
  of the sampled signal. However, these algorithms require either bandlimitedness of the
  signal, or noiseless samples. In this paper, we propose a novel algorithm which does not require the signal to be bandlimited and it can cope with additive noise in the samples.
  The algorithm is illustrated and validated with real data.
\end{abstract}


\section{Introduction}\label{s:intro}

Periodic signal reconstruction from a finite sequence of samples (a sample train) taken at a
given~sampling rate  is an important signal processing
operation needed in many applications in diverse areas such as communications,
remote sensing, system testing, and characterization. If the period $T$ of the
sampled signal is known and the signal is band-limited, then a single sample
train allows for perfect reconstruction of the signal, provided that the train
is long enough \cite{schanze95, candocia98, dooley00}. In such a case, the
reconstruction reduces to solving a set of linear equations. If period $T$ is
not known but the sampling period $\tau$ is much smaller than $T$, then one can
get a reasonable approximation to the sampled signal by interpolation. This is
how the digital scopes usually reconstruct the sampled signals. The situation
becomes much more complicated when $\tau$ and $T$ are of the same order of
magnitude. 
In such a case, we deal with undersampled signals and the methods of
dealing with approximation in this case are often called super-resolution
methods. 
Undersampling might be due purely economic reasons, since
high-speed sampling systems are relatively expensive, or it can result from systems designed or configured
for a low frequency application to recover unanticipated high frequency
signals~\cite{rader77}.
The simplest super-resolution technique is based on the stroboscopic
effect. To take advantage of this effect, one needs to get a~precise estimate
of period~$T$.  This crucial estimation is usually performed in the time domain
\cite{rader77, silva86, bhatta11, tzou12} or in the frequency domain
\cite{choi11}. In both the approaches, a~long train of samples is required to
get a~reasonable period estimate. If the nature of the sampling process or the
sampling hardware  allows for acquisition of short sample-trains only, then one
can still achieve super-resolution by using multiple such trains, even if the
inter-train time distances are not known. This is possible if the sampled
signal is of known band limit~\cite{Vandewalle04, Vandewalle07}, or if the
starting times of the trains are distributed uniformly (in the probabilistic
sense) when considered modulo~$T$~\cite{Rup20icassp, Rup20spl}, or if the
ratio $\tau/T$ is irrational~\cite{Rup20period}. The algorithms for signal
reconstruction proposed by the author in~\cite{Rup20icassp,Rup20spl} require noiseless samples.
In this paper we propose an algorithm which can deal with trains of noisy
samples. The algorithm is based on the ideas introduced in~\cite{Rup20icassp}.

The next section explains the relationship between periodic signals and the
probability distribution of sample trains. Section~\ref{s:algorithm}
proposes an algorithm for reconstruction of periodic signals from a finite
number of trains of noisy samples. 
In Section~\ref{s:experiment}, we present the results of a proof-of-concept experiment which was designed for the proposed algorithm verification. The paper is concluded in Section~\ref{s:conclusion}.

\section{Distribution of trains of samples}\label{s:theory}
\begin{figure}[tp]
  \centering
  \includegraphics{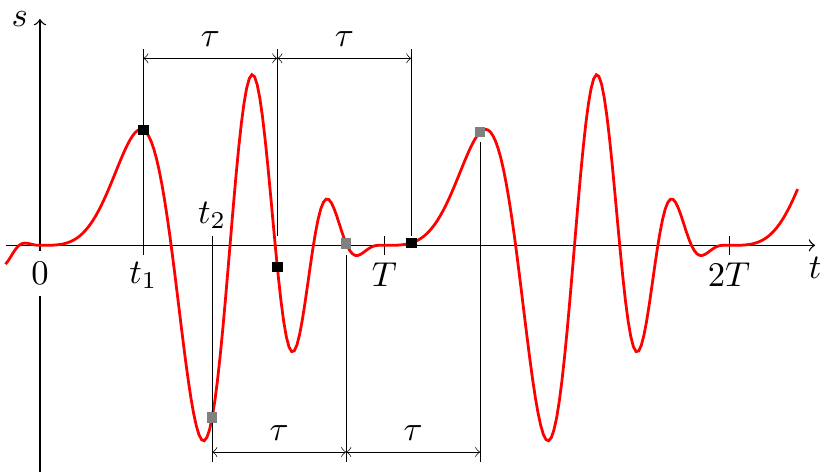}
  \caption{A $T$-periodic signal and its two sample trains of length~$3$ 
    taken with sampling period $\tau\approx 0.4 T$ (the sequences start at
  $t_1$ and $t_2$, respectively) } \label{f:signal1}
\end{figure}
A train of samples of signal $s$ is henceforth denoted by $\samp(t)$, i.e.,
\begin{equation}\label{e:seqdef}
   \samp(t) = [s(t),\, s(t+\tau),\, \dots,\, s(t+(d-1)\tau)]\in\R^d. 
\end{equation}
where $d$ is the length of the train, $t$ is its starting time, and $\tau$ is the
inter-sample distance (sampling period). Fig.~\ref{f:signal1} shows an example of a periodic signal and its 3-sample trains.

If $s$ is a continuous periodic function of variable $t$, then so is the above mapping \samp.
In this case, the image of \samp is a closed curve.
If we treat time instant $t$ in \eqref{e:seqdef} as a random variable
distributed uniformly on the interval $[0,T)$, then the resulting vector
$\samp[d,\tau](t)$ becomes a multivariate \curve-valued random variable
henceforth denoted by~\Sran. In \cite{Rup20icassp} it is shown that the
probability distribution of \Sran determines signal $s$ up to a~time shift
provided that: function \samp restricted to the interval $[0,T)$ is a one-to-one
mapping, the time derivative of $\samp$ does not vanish
anywhere, and $\tau<\frac{T}{2}$.
For the completeness of this paper we rephrase the reconstruction
algorithm presented in~\cite{Rup20icassp} as Algorithm~\ref{a:density}.
There and henceforth, \fc is the probability density function (PDF) of~\Sran, \curve is the closed curve
formed by the trains of samples, i.e., \curve is the image of~function~\samp,
$L$ is the length of~\curve, $\pi_m$ is the projection onto the $m$-th coordinate, i.e., 
$\pi_m([x_1,\dots,x_d]) = x_m$, $f\circ g$ means a composition of the functions, i.e, $f\circ g(t)=f(g(t))$, and $f\transeq g$ means that functions $f$ and $g$ are equal up to a time shift, i.e., there exists $t_0$ such that $f(t)=g(t+t_0)$ for all $t\in \R$.
\begin{algorithm}
    \caption{Signal reconstruction from the probability distribution of its trains of samples (noiseless case) \cite{Rup20icassp}}\label{a:density}
    \textbf{Input:} sampling period $\tau$, PDF $\fc$ and its support \curve\par
    \textbf{Output:} signal $s$ (up to a time shift), and its period $T$.
\begin{enumerate}
    \item Take any arc-length parametrization of curve~\curve and extend it to an $L$-periodic function $\qv_1\colon\R\to\curve$.
    \item \label{step:2} Set $\qv = \qv_1 \circ r$, where function $r\colon\R\to\R$ is given by equation
        $
            x = \int_0^{Lr(x)} \fc(\qv_1(u)) \dd u.
            $
  \item \label{step:3} Take any integers $1\leq k<l\leq d$ and find $x_0\in(0,1)$ such that
      \begin{equation}\label{e:pi1pi2}
          \pi_{k} \circ \qv\left(x + \left(l-k\right)x_0\right) =
          \pi_{l} \circ \qv\left(x\right)\quad \forall x\in\R.
      \end{equation}
    \item \label{step:4} If $x_0 < \frac{1}{2}$, then $T=\frac{\tau}{x_0}$ and 
      $\samp(t) \transeq \qv (\frac{t}{T})$.\par
      If $x_0 >\frac{1}{2}$, then $T=\frac{\tau}{1-x_0}$ and 
      $\samp(t) \transeq \qv (\frac{-t}{T})$.
    \item 
        $ s(t) \transeq \sum_{k=1}^d \pi_k \circ \samp\bigl(t + (k-1)\tau\bigr).$
\end{enumerate}
\end{algorithm}
  
\begin{figure}[tb]
  \centering
  \includegraphics[height=2.1in]{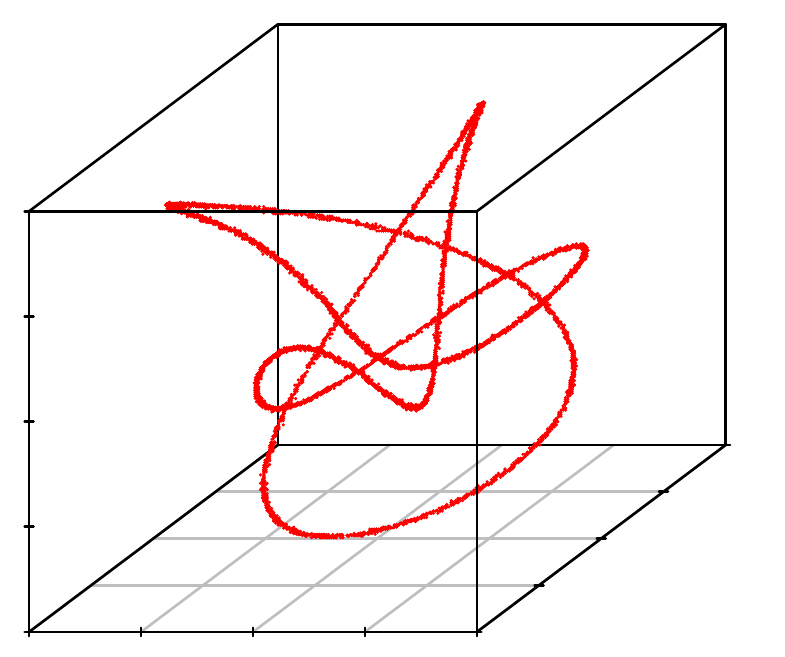}
  \caption{A point cloud formed by a large number of noisy $3$-sample sequences
    of the signal presented in Fig.~\ref{f:signal1} (for bounded noise the
    points lie inside a pipe of a bounded diameter; in the absence of the noise the
    points would form a closed~curve)%
 }\label{f:pcloud}
\end{figure}

We model the additive noise in the samples by assuming that acquired trains of samples are of the form:
\begin{equation}\label{e:noisytrain}
    \samp(t) + \etav \in \R^d,
\end{equation}
where $t$ is a random variable distributed uniformly in the interval $[0,T)$, and $\etav$ is 
a $d$-variate random variable with finite variance $\sigma^2$ and with radial PDF $f_\sigma$, i.e., 
\begin{equation}\label{e:feta}
    f_\sigma(\xv) = c_d\sigma^{-d} f_1\left(\|\xv\|/\sigma\right),
\end{equation}
where $\|\xv\|$ denotes the Euclidean norm of~$\xv$, and where $f_1$ is the PDF of a unit variance positive random variable.
Noisy trains~\eqref{e:noisytrain} no longer lie in a closed curve \curve because the
noise scatters these points around that curve. 
However, if the noise variance is small, the noisy trains fall into a small
neighborhood of~\curve, i.e., curve \curve forms a thread along which the
probability distribution of noisy trains~\eqref{e:noisytrain} is
concentrated.
Fig.~\ref{f:pcloud} illustrates such a~situation by showing a cloud of trains of noisy samples of the periodic signal presented in Fig.~\ref{f:signal1}.

Let $f$ denote the PDF of noisy trains~\eqref{e:noisytrain}. Function $f$ is a convolution of functions $\fc$ and $f_\sigma$, i.e., 
\begin{equation}\label{e:fpv}
    f(\pv) = \int_{\curve}\fc(\qv) f_\sigma(\pv-\qv)\dd \qv.
\end{equation}
Therefore, if $\sigma$ is small, then for each point $\pv\in\curve$
\begin{equation}\label{e:fpapprox}
    f(\pv) \approx c_d\sigma^{-(d-1)} \fc(\pv).
\end{equation}
Indeed, we can split the integral in~\eqref{e:fpv} into two parts: the integral over a $\sqrt{\sigma}$-neighborhood of \pv, and over the rest of curve~\curve, respectively.
When $\sigma$ tends to $0$, the former integral becomes the right hand side
of~\eqref{e:fpapprox}, while the latter integral is bounded by $\sigma^{-(d-1)}h(\sigma)$, where $h(\sigma)$ is proportional to $\int_{1/\sqrt{\sigma}}^\infty f_1(y)\dd y \stackrel{\sigma\to 0}{\to} 0$.
Thus, the smaller the variance of $\etav$ is, the more accurate approximation~\eqref{e:fpapprox} is.

\section{The reconstruction algorithm}\label{s:algorithm}
The trains of samples used for the reconstruction
of a signal are henceforth treated as points in a $d$-dimensional space and they are denoted by 
\begin{equation}\label{e:pv}
    \pv_1,\dots,\pv_n\in\R^d.
\end{equation}
The outline of the proposed reconstruction algorithm is presented below as Algorithm~\ref{a:outline}. The following subsections explain 
in detail the three stages of the outline.
\begin{algorithm}[H]
    \caption{Signal reconstruction from a finite number of trains of noisy samples (outline)}\label{a:outline}
    \textbf{Input:} sampling period $\tau$, trains of samples~\eqref{e:pv}\par
    \textbf{Output:} estimates $\est{s}$ and $\est{T}$ of signal $s$ and its period $T$, resp.
    \begin{enumerate}
        \item Approximate points~\eqref{e:pv} with a closed smooth curve \est{\curve},
        \item Estimate a PDF $\est{\fc}$ along curve \est{\curve} with a help of~\eqref{e:fpapprox},
        \item Compute \est{s} and \est{T} as the output of appropriately adapted Algorithm~\ref{a:density} applied to: $\tau$, \est{\fc}, \est{\curve} 
    \end{enumerate}
\end{algorithm}
\subsection{Curve \curve approximation}
The problem of recovering a curve from its noisy samples (cloud of points)
appears in several applications such as Computed Axial Tomography,
Coordinate-Measuring Machine measurements and Magnetic Resonance Imaging. There
exist various strategies to solve this problem.  They
are based on methods of mathematical analysis~\cite{fang1995, goshtasby2000},
geometry~\cite{pottmann1998, cheng2005} and statistics~\cite{hastie1989, levin1998,lee2000, Rup14}.  
The approximation algorithm presented in~\cite{Rup14} is especially well suited for the first stage of Algorithm~\ref{a:outline} because
it can cope with closed curves, it works in Euclidean space
of arbitrary dimension, and it is relatively simple. The algorithm of~\cite{Rup14} has a~parameter
$R$ which can be adjusted for a given or expected noise variance.
The output of the algorithm of~\cite{Rup14} comprises sequences of points, where each sequence represents the
nodes of a polygonal chain which approximate a~curve. If the first and the last
points of an output sequence coincide, then the corresponding polygonal chain
represents a closed curve.
An output sequence which does not represent a closed curve, as well as more than one
output sequences, signals that either the algorithm parameter $R$ is too big, or
the number of points $n$ is not big enough, or eventually that the underlying curve
has some self-intersections. 
If the algorithm of~\cite{Rup14} produces only one output sequence
\begin{equation}\label{e:tildepv}
    \tilde{\pv}_1,\dots,\tilde{\pv}_M\in\R^d,
\end{equation}
and if this sequence represents a closed polygonal path, i.e.,
$\tilde{\pv}_M=\tilde{\pv}_1$, then what remains to complete the first stage of
Algorithm~\ref{a:outline} is to interpolate the nodes of the path with a~smooth
curve. This can be achieved with periodic cubic splines~\cite{Bartels87}. For
the needs of the next subsection, we
denote the resulting smooth curve parametrization by
$\est{\qv}_1\colon[0,\est{L})\to\R^d$. Without loss of generality, we henceforth assume that
$\est{\qv}_1(0) = \tilde{\pv}_1$ and that $\est{\qv}_1$ is already an arc-length
parametrization, i.e. $\|\est{\qv}'_1\|=1$ and $\est{L}$ is the length of curve~\est{\curve} which is the image of~$\est{\qv}_1$. 

\subsection{Density estimation}
Let $\tilde{t}_1,\dots,\tilde{t}_M$ denote the preimages of points~\eqref{e:tildepv} under $\est{\qv}_1$, i.e., 
\begin{equation}\label{e:tti}
    \est{\qv}_1(\tilde{t}_i) = \tilde{\pv}_i,\quad i=1,\dots,M.
\end{equation}
In order to estimate density along curve~\est{\curve}, for each point
$\tilde{t}_i$ we compute the number $c_i$ of
points~\eqref{e:pv} that lie in the $R$-neighborhood of
$\tilde{\pv}_i$, where $R$ is the same as
described in the previous subsection. We obtain \est{\fc}
by linear interpolation of values $c_i$ followed by division of the resulting
function by a factor which makes the interpolation function a~PDF, i.e., by a factor which makes the
integral of the so obtained estimate $\est{\fc}$ unitary. This factor can be expressed as
\begin{equation}
    \sum_{i=1}^{M-1} \left(\tilde{t}_{i+1}-\tilde{t}_i\right)\frac{c_i+c_{i+1}}{2}.
\end{equation}

\subsection{Signal reconstruction}
In the final stage of Algorithm~\ref{a:outline} we refer to
Algorithm~\ref{a:density}.  However, Algorithm~\ref{a:density} has to be
adapted for the fact that $\est{\fc}$ and $\est{\qv}$ are estimates only and they could make
Equations~\eqref{e:pi1pi2} not solvable for any $1\leq k<l\leq d$. 
Therefore, we look for a~least-squares solutions to the set of Equations~\eqref{e:pi1pi2}, i.e., we find $x_0$ by minimizing a functional $F$ in the
interval $(0,1)$, where
\begin{multline}\label{e:min}
    F(x_0) =\\
    \sum_{1\leq k<l\leq d}
    \int_0^1 \left(
        \pi_k \left( \qv\left(x+(l\!-\!k)x_0\right)\right) -
        \pi_l \left( \qv\left(x\right)\right)
    \right)^2\dd x.
\end{multline}

\section{Proof of concept experiments}\label{s:experiment}
\begin{figure}[tbh]
  \centering
  \includegraphics{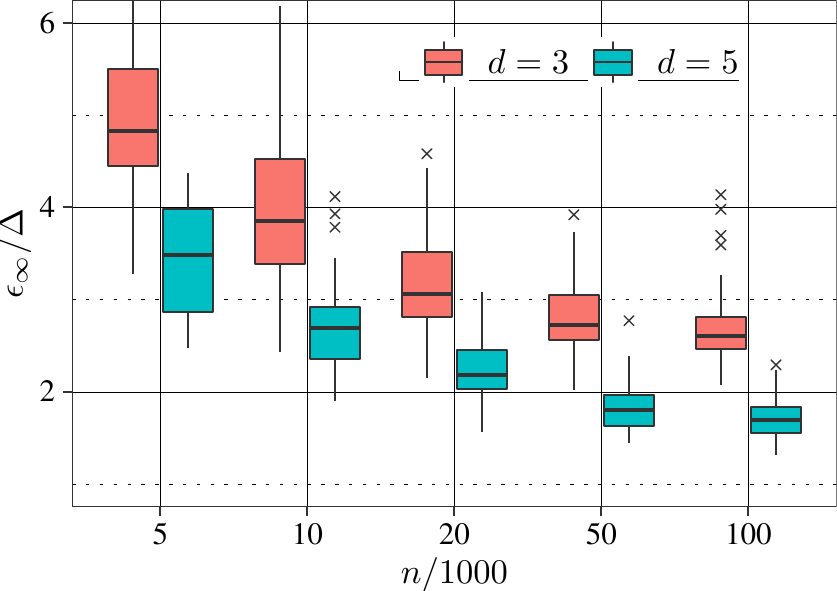}
  \caption{Signal reconstruction maximum errors~\eqref{e:bladinf}} \label{f:boxI}
\end{figure}

\begin{figure}[tbh]
  \centering
  \includegraphics{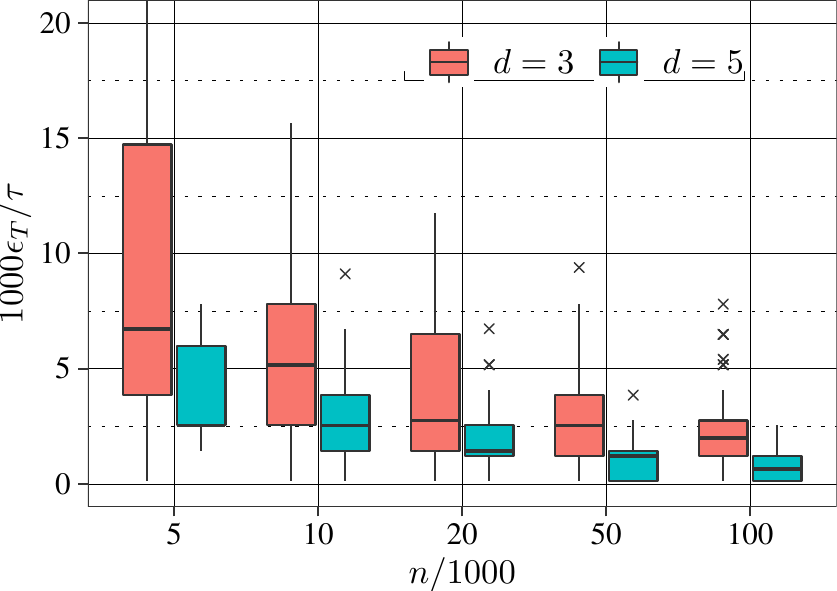}
  \caption{Period estimation errors~\eqref{e:bladT}} \label{f:boxT}
\end{figure}

To verify the proposed algorithm, we have designed the following experiments.
A chirp-like $T$-periodic signal presented in Fig.~\ref{f:signal1} was generated with
a function generator and recorded with a
digital scope. The sampling period $\tau$ of the scope was set to around
$0.39T$. The signal peak-to-peak value was
around \SI{4}{V} and the quantization step of the scope was
$\Delta=\SI{0.02}{V}$ (we may assume that the noise standard deviation $\sigma$ is of the same order of magnitude).  For a fixed value of sample train length $d$ and a fixed
number $n$, we picked at random $n$ $d$-sample trains from the long sample
sequence recorded with the scope. These sequences, treated
as points \eqref{e:pv}, and the sampling period $\tau$ made up
the input data for Algorithm~\ref{a:outline}, in which we set $R=5\Delta$ (the
parameter is needed at the first two stages of the algorithm). The reconstructed
signal period $\est{T}$ and signal $\est{s}$ were then compared to known period $T$ and a~reference signal $s_\text{ref}$, respectively.
The reference signal was obtained by recording the averaged (by the scope) signal with the same digital scope, but with the sampling
period set to $\tau_{\text{ref}} < \tau/1000$. 
\begin{figure}[th]
  \centering
  \includegraphics{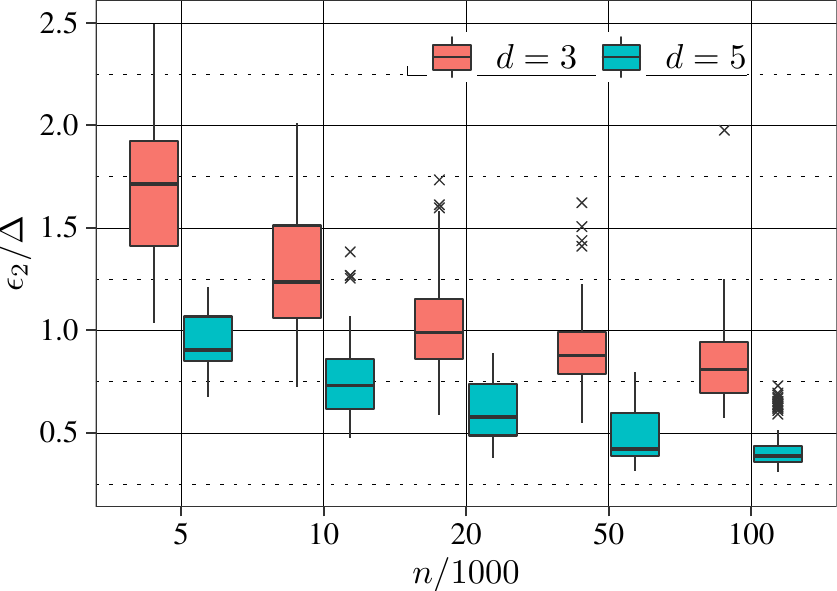}
  \caption{Signal reconstruction root-mean-square errors~\eqref{e:blad2}}\label{f:box2}
\end{figure}
To assess the quality of reconstruction, the following reconstruction error measures
were computed:
\begin{align}
    \blad_{T} &= |T-\est{T}|,\label{e:bladT}\\
  \blad_{2} &= \min_{t_0\in[0,T)} \sqrt{ \int_{0}^T \left(s_\text{ref}(t-t_0) -
  \est{s}(t\est{T}/T)\right)^2},\label{e:blad2}\\
          \blad_{\infty} &= \min_{t_0\in[0,T)} \max_{t\in[0,T)} |s_\text{ref}(t-t_0) - \est{s}(t \est{T}/T)|.\label{e:bladinf}
\end{align}
(the argument of the reconstructed signal \est{s} in~\eqref{e:blad2} and~\eqref{e:bladinf} is scaled to compare two periodic signals of the same period).
For each pair of the considered values of parameters $d$ and $n$, the
experiment was repeated $100$ times.  
The resulting errors are presented as box-plots in
Fig.~\ref{f:box2}--\ref{f:boxT}. The figures show that for the signal considered in the experiment, the super-resolution signal reconstruction has been achieved.
The period estimate error is by three order of magnitude lower than the sampling period~$\tau$.
The both root-mean-square and maximum errors of signal reconstructions reach the level of the quantization step $\Delta$ of the recording of the reference signal.
Not surprisingly, the experiment showed that the quality of signal
reconstruction rises with the amount of input data. The reason for this is
two-fold.  First, the number of cloud points affects the quality of curve
reconstruction from that cloud (the first stage of the algorithm). A
quantitative study of this phenomenon and the role of the dimension of the
space and the amplitude of the noise in the process of curve reconstruction
from a point cloud can be found in~\cite{Rup14}. Second, the amount of input
data affects the density estimation along the reconstructed curve. We expect
that the root-mean-square error of such estimation is similar to that based on histogram, i.e.,
it falls at rate $n^{-1/3}$~\cite{Wasserman10}.


\section{Conclusion}\label{s:conclusion}
  We have proposed an algorithm for signal reconstruction from the short trains
  of noisy samples of the signal. 
  The algorithm allows for achieving a~super-resolution reconstruction as it allows the sampling period $\tau$ to be relatively big with respect to the signal period~$T$
  (the algorithm requires only $\tau<\frac{1}{2}T$).
  We performed a series of experiments, which showed that the reconstructed signal is close to the reference signal and that the larger the number of trains is, the smaller reconstruction errors are obtained.
  We believe that the ability of the algorithm
  to cope with noisy and non-continuous recordings (multiple trains of samples) will make it
  a valuable tool in a number of applications.
  


\bibliographystyle{IEEEbib}
\bibliography{IEEEabrv,biblio}
\end{document}

%% file: skroty.tex
\usepackage{xspace}
\newcommand{\samp}[1][d,\tau]{\ensuremath{\mathbf{s}_{#1}}\xspace}
\newcommand{\Sran}[1][d,\tau]{\ensuremath{\mathcal{S}_{#1}}\xspace}
\newcommand{\R}{\ensuremath{\mathbf{R}}\xspace}
\newcommand{\curve}{\ensuremath{\mathcal C}\xspace}

\newcommand{\fc}{\ensuremath{f_{\curve}}\xspace}

\newcommand{\est}[1]{\ensuremath{\hat{#1}}\xspace}
\newcommand{\vect}[1]{\ensuremath{\mathbf{#1}}\xspace}

\newcommand{\xv}{\vect{x}}
\newcommand{\pv}{\vect{p}}

\newcommand{\qv}{\vect{q}}
\newcommand{\etav}{\vect{\eta}}

\newcommand{\dd}{\ensuremath{\mathrm{d}}}

\newcommand{\trans}[1][]{\ensuremath{\phi_{#1}}\xspace}
\newcommand{\transeq}{\ensuremath{\stackrel{\trans}}{=}}

\newcommand{\blad}{\ensuremath{\epsilon}}